\documentstyle[12pt]{article}
\begin{document}     

\begin{center}         
{ \Large   CORPUSCULAR INTERPRETATION OF MIKROPARTICLE
DIFFRACTION AND POSSIBLE  EXPERIMENTAL  TEST}

\vskip 1cm
 G.L.Bayatian   

\vskip 1cm
{\it Yerevan Physics Institute,\\ Yerevan 375036,Armenia }              
\end{center}

\abstract{     
     We examine the logical difficulties that arise when a particle(wave)
 interferes with itself.We propose to carry out to its full extent a
 ,gedanken' experiment originally proposed by Feynman in order to give an
 unequivocal experimental proof of the correctness of the wave
 representation.As an example,we give a corpuscular interpretation of the 
diffraction pattern.
}        
               
\section{Introduction}
     
     Wave interpretation of quantum phenomena was started in 1924 by
 de Broglie hypotesis,which says that wavelength of any particle with P
 momentum is equal to
\begin{equation}
 \lambda = h/P   
\label{1}
\end{equation}

 where $h$ is Plank's constant.

     This hypothesis was first proved by experiment conducted by K.Davisson
 and L.Djermer \cite{davisson}. In this experiment by observing reflection of electron
 beam from nickel single cristal they have obtained pattern absolutely
 corresponding to wave diffraction with $\lambda$, defined by (\ref{1}) and satisfied by
 the condition
\begin{equation}
b Sin(\theta) = n \lambda         
\label{2}
\end{equation}
 where $b$ is cristal period,n is whole number,$\theta$ is observation angle of
 diffraction maxima.

     To develop a stable (observable) diffraction pattern in classic optics
 it is required to have simultaneous arrival to point of observation of 2
 (interference) or more than 2 (diffraction) monocromatic and coherent waves.
  From the point of view of quantum mechanics, one will obtain interference of
 wave with itself if a barrier with 2 slits will be put on a way of single
 particle.By multiple repetition of the experiment full diffraction picture
 will be received on the screen.The probable nature of wave function F allows
 doing that (of course, on paper).

       Let's assume that a microparticle passes through the first slit with
 probability $\omega_{1}=|\psi_{1}|^{2}$, and through the second slit with 
$\omega_{2}=|\psi_{2}|^{2}$.The
 superposition principle allows to write the following for the wave
 function $\Psi$
\begin{equation}     
\Psi  = c_{1} \psi_{1} + c_{2} \psi_{2},   
\label{3}
\end{equation}
 where $c_{1}$ and $c_{2}$ are normalization coefficients. By choosing 
$\psi_{1}=a_{1}Sin(\omega t-\phi_{1})$ and $\psi_{2}=a_{2}Sin(\omega t-\phi_{2})$
 we will have
\begin{equation}         
W=| \Psi |^{2}=a_{1}^{2}+a_{2}^{2}+2a_{1}a_{2}Cos(\phi_{1}-\phi_{2}),   
\label{4}
\end{equation}
 where $W$ is probability of particle to hit certain point of screen.

    One can see from (\ref{4}) that depending on $(\phi_{1}-\phi_{2})$ the 
last (interference)
 term can be either positive or negative,i.e.in different points of the
 screen the hitting probability of particles will be different,and by long
 exposure of the screen by single particles the full diffraction picture can
 be received.
    Let's see what logical contradictions will occur from superposition
principle in interpretation of the diffraction pattern.
    No doubt that by passing a barrier with two slits the particle(wave) is not
 spliting into 2 parts, otherwise it would be visible by experiment.This means 
that each particle passes through one of these two slits(doesn`t matter through
 wich one exactly).Then in expression (\ref{3}) for wave function $\Psi$ beyond the 
barrier either $\psi_{1}=1$, and consequently $\psi_{2}=0$ , or $\psi_{2}=1$, and 
$\psi_{1}=0$. In both cases the
 interference term disappears in (\ref{4}).The interference maxima and minima 
observed in all experiments show that wave formalism introduced to explain
 diffraction pattern contradicts to experimental results.This conclusion
becomes obvious if one looks at (\ref{3}) from the point of view of simultaneous 
arrival of two waves to point of observation.To satisfy this condition ,
obligatory for the interference,the quantum mechanics divides tasitly the
 particle (wave) into 2 parts with the help of equition(3).Just in the same 
tacit way it is assumed that passing various paths these divided parts meet each
 other at a microscopic area $\lambda^{2}=10^{-16}sm^{2}$ ($\lambda$  is the de 
Broglie wavelengt equal
 to about $1A^{o}$ in the experiments \cite{davisson} and \cite{carnal}) of the 
macroscopic screen surface $\sim 10 cm^{2}$.
 
   In this work we show that the observed maxima and minima of the diffraction
 pattern can have other origin ,connected with the action discretness which
 served for the development of the quantum mechanics.In this way one can escape
 the wave presentation and the connected with them logical difficulties of 
 explaining the interference of single particle(wave) with itself. To obtain an
 answer to this important question we propose a ,gedanken' experiment considered
 by Feynman,which will give unambigeous answer on the validity of the wave
 presentations.
    The discussion of the corpuscular interpretation of the diffraction pattern
 is also given.

\section{Two slits  experiment}
        
    According to \cite{cern} the attempt to understand the phenomenon of single particle
 (wave) interference with itself is an unnecessary intellectual masochism.
 However,the absence of the unambigous test of the influence of the second slit
 through which the particle did not pass, makes such a statement groundless.
    The contemporary atomic interferometers \cite{carnal,keith,chapman} allow one to 
realize such a
 test according to the folloving scheme:it is necessary to compare the 
 interferometer pattern, obtained with the help of two slit interferometer \cite{carnal},
 for  instance, when both slits are open,with the summary pattern composed of two
 expositions when consequently one of the slits is closed. In the last case the
 interference term in (\ref{4}) is absent and the maxima and minima must not be
 observed. If the summary pattern will differ from the one when both slits are
 open,then this will be anambigeous experimental confirmation of the fact that
 the open slit, through  which the particle did not pass,is not equivalent to
 the close slit and it influences the interference pattern. And vice versa,if
 the summary pattern will be identical to the one when both slits are open,then
 this will mean that the wave interpretation of the appearance of maxima and 
 minima is incorrect,and it is necessary to look for other explanation.

\section{The corpuscular interpretation}
 
    To overcome the above mentioned well known logical contradictions, connected
 with the interference of a single particle (wave) with itself,it is necessary
 to take into account the interaction of this particle with the slit matter
 (diffraction grating) as well as with subsidiary exiting electron \cite{carnal} and
 laser \cite{chapman} beams. At the first sight,the problem seems to have no solution.
 However, it can be solved using Planck's constant $h$, common for all types of 
 interactions quantum of action,together with the hypothesis that the action is
 multiple to $h$ for the unbound states,as for bound states. Let us follow how
 this can be done.

      Let a parallel beam of microparticles wit momentum $P$ moves along the
 direction $x$ and falls on a single cristal with period $b$ (distance between the 
 cristallographic planes),or on a slit with width $b$. As a result of interaction
 with the cristal(slit) matter the particle gets a transverse momentum $P_{r}$ and 
 will be scattered under an angle $\theta$, defined by the relation:
\begin{equation}                 
Sin(\theta ) = P_{r}/P  
\label{5}
\end{equation}
It is necessary to find $P_{r}$ from the following differential equation
\begin{equation}          
dP_{r} = F_{r} dt  
\label{6}
\end{equation}                               
 where $F_{r}$ is the force acting on the particle by the cristal in a direction
 perpendicular to the beam direction.
      
Multiplying (\ref{6}) by $dr$ one obtains
\begin{equation}       
             dP_{r} dr= F_{r} dr dt = dS_{r}
\label{7}
\end{equation}
 where $dS_{r}$ is the action on the path $dr$ in the direction $r$ for the time $dt$.
 Using the hypothesis of multiplicity of action to $h$ and choosing the 
 integration limits  from 0 to $t$ for the time and from 0 to $b$ for $r$ one obtains
\begin{equation}
               b P_{r} = S_{r} = n h ,
\label{8}
\end{equation}
 or taking into account (\ref{5})
\begin{equation}              
               b Sin (\theta ) = n h/P = n \lambda 
\label{9}
\end{equation}
 where $\lambda$ is the well known de Broglie wave length.
     Formula (\ref{9}) shows that the agreement the experimental data with the
 de Broglie hypothesis is not accidental. In a hidden form the de Broglie
 hypothesis contains the hypothesis of the multiplicity of action to $h$ in an
 interaction used in this work. Just for this reason,as it is seen from (\ref{9})
 the diffractive scattering angles take discrete  volues,imitating the 
 diffraction pattern.
            
\section{Conclusion}

     To test the above proposed interpretation it is very important to perform
 the Feynmans,gedanken' experiment with two slits completely with the help of
 atomic interferometers.


\begin{thebibliography}{99} 
\bibitem{davisson} C.Davisson and L.Germer,Phys.Rev.30,705 (1927).         
\bibitem{carnal}   O.Carnal and J.Mlynek,Phys.Rev.Lett.66,2689(1991).
\bibitem{cern}     CERN COURIER,v.36,n.2 ,p.27 (1996).
\bibitem{keith}    D.W.Keith,C.R.Ekstrom et al.,Phys.Rev.Lett.66,2693(1991).
\bibitem{chapman}  M.S.Chapman et al.,Phys.Rev.Lett.75,3783(1995).
\end{thebibliography}
 \end{document}